\documentstyle[aps,prl,12pt]{revtex}
\begin{document}

\preprint{\vbox{\baselineskip=3ex
                \hbox{KEK-TH-493}
                \hbox{KEK preprint 96-111}
                \hbox{September 1996}
                \hbox{H}}
                \bigskip}
\draft

\title{Background Suppression for $\mu\rightarrow e\gamma$ with
Polarized Muons}

\author{Yoshitaka Kuno, Akihiro Maki, and Yasuhiro Okada}

\address{Department of Physics, National Laboratory for High Energy
Physics (KEK),\\ Tsukuba, Ibaraki, Japan 305}

%\date{\today}

\maketitle

\begin{abstract}

A search for the lepton-flavor violating $\mu^{+}\rightarrow
e^{+}\gamma$ decay using polarized muons is revisited in terms of
suppression of the serious background arising from an accidental
coincidence between a $e^{+}$ in the normal muon decay and a high
energy photon in the radiative muon decay, $\mu^{+} \rightarrow e^{+}
\nu \overline{\nu} \gamma$.  It is found that a high energy photon in
$\mu^{+} \rightarrow e^{+} \nu \overline{\nu} \gamma$ decay is
preferentially emitted parallel to the muon spin direction, similarly
to $e^{+}$ in the normal muon decay. The selective measurements of
either $e^{+}$s or photons moving antiparallel to the muon spin
direction would suppress the accidental background for $\mu^{+}
\rightarrow e^{+} \gamma$ with right-handed and left-handed $e^{+}$s
respectively.

\end{abstract}

\bigskip


%\narrowtext
%\begin{multicols}{2}
%\parskip 0pt

A search for lepton-flavor violating (LFV) processes such as $\mu^{+}
\rightarrow e^{+} \gamma$ decay and $\mu^{-}-e^{-}$ conversion in a
nucleus has been attracting much theoretical and experimental
interest, since various theoretical models with physics beyond the
Standard Model predict a large LFV effect. In particular,
supersymmetric grand unification (SUSY-GUT) predicts large branching
ratios for those LFV processes, which are only one or two orders of
magnitude lower than the current experimental limits
\cite{hall86}. Some other theoretical extensions to the Standard Model
also predict a large branching ratio \cite{hisa95,verg86}.

The event signature of $\mu^{+} \rightarrow e^{+} \gamma$ is that a
$e^{+}$ and a photon are in coincidence, and moving colinear
back-to-back with their energies equal to a half of the muon mass
($m_{\mu}/2 = 52.8$ MeV).  The current experimental upper limit is
$4.9 \times 10^{-11}$ at 90 \% confidence level \cite{bolt88}.

One of the major background to the search for $\mu^{+} \rightarrow
e^{+} \gamma$ is a radiative muon decay $\mu^{+} \rightarrow e^{+} \nu
\overline{\nu} \gamma$ (branching ratio is 1.4 \% for $E_{\gamma} > 10$
MeV). When $e^{+}$ and photon are emitted back-to-back with two
neutrinos carrying off little energy, it becomes a serious physics
background to $\mu^{+} \rightarrow e^{+} \gamma$.  The other
background, which turns out more important in a new generation
experiment with a very high rate of stopped muons, is an accidental
coincidence of a $e^{+}$ in a normal muon decay, $\mu^{+} \rightarrow
e^{+} \nu \overline{\nu}$, accompanied by a high energy photon. The
sources of a high energy photon might be either that in $\mu^{+}
\rightarrow e^{+} \nu \overline{\nu} \gamma$ decay, or external
bremsstrahlung or annihilation-in-flight of $e^{+}$s in the normal
muon decay.

In our previous paper by two of the authors\cite{kuno96}, the
importance has been emphasized of the use of polarized muons in a
search for $\mu^{+} \rightarrow e^{+} \gamma$. With the use of
polarized muons, a right-handed $e^{+}$ ($e^{+}_{R}$) in $\mu^{+}
\rightarrow e^{+}_{R} \gamma$ decay follows a $1 -
P_{\mu}\cos\theta_{e}$ distribution, whereas a left-handed $e^{+}$
($e^{+}_{L}$) in $\mu^{+} \rightarrow e^{+}_{L} \gamma$ does a $1 +
P_{\mu} \cos\theta_{e}$ distribution, where $\theta_{e}$ is an angle
of the $e^{+}$ emission with respect to the muon polarization
($P_{\mu}$). This angular distribution is useful to discriminate
different models. More importantly, it is shown that the physics
background from the radiative muon decay has a $e^{+}$ following
approximately a $1 + P_{\mu} \cos\theta_{e}$ distribution when the
energy resolution of photon detection is worse than that of
$e^{+}$. Also for the accidental background, $e^{+}$ in the normal
muon decay is known to follow a $1+P_{\mu}\cos\theta_{e}$
distribution. They imply that the selective measurement of $e^{+}$s
antiparallel to the muon polarization direction would suppress these
backgrounds, improving a sensitivity of the search only for $\mu^{+}
\rightarrow e^{+}_{R} \gamma$. No background suppression, however, is
expected for $\mu^{+} \rightarrow e^{+}_{L} \gamma$.

In this Letter, we present our further studies on the accidental
backgrounds with polarized muons. In particular, in order to
investigate the accidental background for $\mu^{+} \rightarrow
e^{+}_{L} \gamma$ decay, we examine the angular distribution of a
high energy photon from $\mu^{+} \rightarrow e^{+} \nu \overline{\nu}
\gamma$. It is found to be emitted preferentially along the muon spin
direction; namely follows a $1 + P_{\mu} \cos\theta_{\gamma}$
distribution, where $\theta_{\gamma}$ is an angle of the photon
direction with respect to the muon spin direction. It implies that the
accidental background could be suppressed for $\mu^{+} \rightarrow
e^{+}_{L} \gamma$ where high energy photons must be detected at the
opposite direction to the muon polarization. This is the same
suppression mechanism seen for $\mu^{+} \rightarrow e^{+}_{R} \gamma$
where high energy $e^{+}$s going antiparallel to the muon polarization
direction are measured. As a result, the selective measurements of
either $e^{+}$s or photons antiparallel to the muon spin direction
would give the same accidental background rejection for $\mu^{+}
\rightarrow e^{+}_{R} \gamma$ and $\mu^{+} \rightarrow e^{+}_{L}
\gamma$ decays respectively. 

To examine the angular distribution of a high energy photon in
$\mu^{+} \rightarrow e^{+} \nu \overline{\nu} \gamma$ decay, we
calculated the differential decay width of the radiative muon decay,
retaining the positron mass ($m_e$), as a function of $e^{+}$ energy
($E_{e}$) and photon energy ($E_{\gamma}$) \cite{okad96}. Those
energies are normalized to a half of the muon mass ($m_{\mu}/2$),
namely $x = 2E_{e}/m_{\mu}$ and $y = 2E_{\gamma}/m_{\mu}$. From the
four-body kinematics, the ranges of $x$ and $y$ are given as follows;
$y$ varies from 0 to $1-r$ where $r = (m_{e}/m_{\mu})^2$.  For $0 < y
\leq 1-\sqrt{r}$, $x$ changes from $2\sqrt{r}$ ({\it i.e.} the
normalized positron rest mass) to $1+r$, and for $1 - \sqrt{r} < y
\leq 1 - r$, $(1-y) + r/(1-y) \leq x \leq 1+r$. The differential decay
width thus obtained is identical to that in Reference \cite{fron59}
when the positron mass is ignored. Since the $e^{+}$s in the radiative
muon decay do not contribute to the accidental background, the
differential decay width should be integrated over the $e^{+}$ energy
and angle between $e^{+}$ and photon ($\theta_{e\gamma}$) in the
kinematically allowed region. Only high energy photons in the extreme
kinematic case of $y \approx 1$ can slip into the signal
region. Taking these into account, in the limit of $y \approx 1$, the
differential branching ratio is approximately given by

\begin{equation}
dB(\mu^{+} \rightarrow e^{+} \nu \overline{\nu} \gamma) =
\Bigl[ J(y)(1+P_{\mu}\cos\theta_{\gamma})+ O((1-y)^2, \sqrt{r}) \Bigr] dy
d(\cos\theta_{\gamma} ) 
\label{eq:ph50}
\end{equation}

\noindent where $J(y)$ is given by

\begin{equation}
J(y) =  {\alpha \over 2\pi}(1-y)\Bigl[ {\rm ln}{(1-y)\over r} -
{17\over6} \Bigr]
\label{eq:oka}
\end{equation}

\noindent As seen in Eq.(\ref{eq:ph50}), the angular distribution of
high energy photon ($y \approx 1$) follows a 1 + $P_{\mu}
\cos\theta_{\gamma}$ distribution when the higher order terms of
$(1-y)$ are neglected.  Equation (\ref{eq:oka}) is derived under the
assumption of $2\sqrt{r} \ll 1-y \ll 1$. It should be noted that
Eq.(\ref{eq:ph50}) is consistent with the other calculations for the
case of unpolarized muons \cite{kino59}. A rate of the observed events
in a real experiment can be estimated by integrating the spectrum over
the photon energy resolution of the detector, or more precisely the
width of the signal region. Taking $\delta y$ to be a half width of
the signal region for photon energy, the partial branching ratio
(denoted by $b_{\gamma}$) integrated over the signal region ($1-\delta
y \leq y \leq 1-r$) can be calculated from Eq.(\ref{eq:ph50}) by

\begin{eqnarray}
b_{\gamma} &=& \int^{1-r}_{1-\delta y}dy 
{dB(\mu^{+}\rightarrow e^{+}\nu\overline{\nu}\gamma) \over dy} \cr
&\approx& \Bigl({\alpha\over4\pi}\Bigr) (\delta y)^2 
\Bigl[\ln(\delta y) + 7.33 \Bigr](1 + P_{\mu} \cos\theta_{\gamma})
d(\cos\theta_{\gamma})
\label{eq:int50}
\end{eqnarray}

\noindent From Eq.(\ref{eq:int50}), it is shown that $b_{\gamma}$ is
roughly proportional to $(\delta y)^2$. 

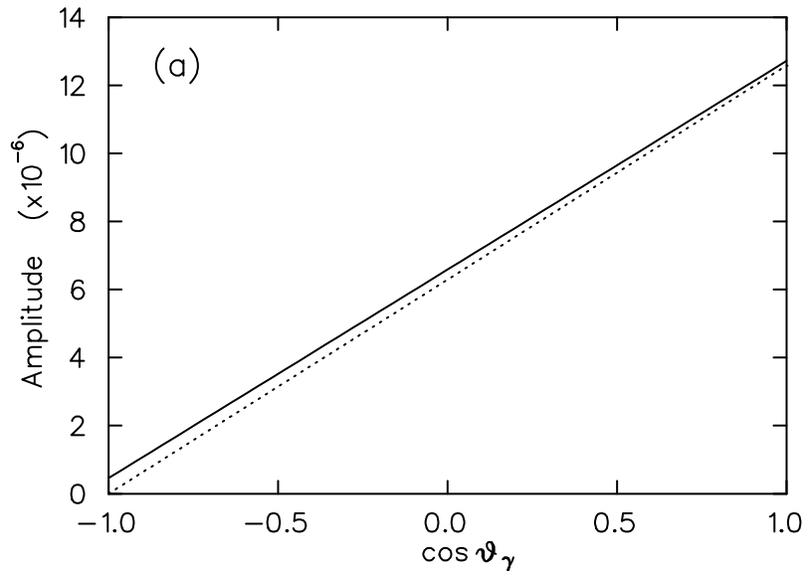
\begin{figure}
\caption{Angular dependence of partial decay width of $\mu^{+}
\rightarrow e^{+} \nu \overline{\nu} \gamma$ as a function of cosine
of an angle between the photon and muon spin directions. Solid lines
are the numerical calculations. Dotted lines are from Eq.(3). (a)
$\delta y$ = 0.05, (b) $\delta y = 0.03$, (c) $\delta y = 0.01$.}
\label{fg:ph50}
\end{figure}

The numerical calculation of $b_{\gamma}$ was also carried out based
on the exact expression of the differential decay width. The effect of
the positron mass was taken into account in the evaluation of
kinematic boundaries of $x$ and $\theta_{e\gamma}$, as well as in the
expression of the differential decay width, although the latter turned
out to be small, about a few \%. In Fig.\ref{fg:ph50} the numerically
calculated $b_{\gamma}$ is shown as a function of
$\cos\theta_{\gamma}$ for various values of $\delta y$, 0.05, 0.03,
and 0.01. Fig.\ref{fg:ph50} also shows $b_{\gamma}$ given in
Eq.(\ref{eq:int50}), which is in agreement with the numerically
calculated $b_{\gamma}$. As seen in Fig. \ref{fg:ph50}, the higher the
energy of detected photon is ($i.e.$ $\delta y$ becomes smaller), the
closer to the $1 + P_{\mu}\cos\theta_{\gamma}$ the angular
distribution of $b_{\gamma}$ becomes.  For $\delta y = 0.03$ with
$P_{\mu}$ = 100 \%, the constant term remaining at
$\cos\theta_{\gamma} = -1$ is about 2.1 \% of the value at
$\cos\theta_{\gamma} = +1$.

The other sources of high energy photons ($y \approx 1$) are external
bremsstrahlung and annihilation-in-flight of $e^{+}$s in the normal
muon decay. Since $e^{+}$s in the normal muon decay is known to follow
a $1+P_{\mu}\cos\theta_{e}$ distribution, photons from these sources
are also emitted preferentially along the muon spin direction. It is
noted that they could be minimized experimentally by reducing
materials around the target, and are found in previous experiments to
be less significant than that from the radiative decay \cite{kinn82}.

From these, it is concluded that both of the two major sources of
accidental background, high energy photons ($y \approx 1$) from the
radiative muon decay and high energy $e^{+}$ ($x \approx 1$) from the
normal muon decay, follow $1 + P_{\mu}\cos\theta$ distribution, where
$\theta$ is either $\theta_{e}$ or $\theta_{\gamma}$. In a search for
$\mu^{+} \rightarrow e^{+}_{R} \gamma$ where a $e^{+}$ is emitted
antiparallel to the muon spin direction, the suppression of accidental
background results from the selective measurement of $e^{+}$ moving
antiparallel to the muon spin. In a search for $\mu^{+} \rightarrow
e^{+}_{L} \gamma$ where a high energy photon is antiparallel to the
muon spin direction, the same is the case if a $e^{+}$ is replaced
with a photon. As a result, the accidental background could be
suppressed for both $\mu^{+} \rightarrow e^{+}_{R} \gamma$ and
$\mu^{+} \rightarrow e^{+}_{L} \gamma$ decays by the use of polarized
muons.

A quantitative discussion of the suppression factor is given below.
The rate of the accidental background ($B_{acc}$) normalized to the
total decay rate can be estimated by

\begin{equation}
B_{acc} = R_{\mu} 
\cdot f^{0}_{e} \cdot f^{0}_{\gamma} 
\cdot (\Delta t) \cdot ({\Delta \omega \over 4\pi}) \cdot \eta
\end{equation}

\noindent 
where $R_{\mu}$ is an instantaneous muon intensity.  $f^{0}_{e}$ and
$f^{0}_{\gamma}$ are defined in $f_{e} \equiv f^{0}_{e}(1 +
P_{\mu}\cos\theta_{e}) (d\Omega_{e}/4\pi)$ and $f_{\gamma} \equiv
f^{0}_{\gamma}(1 + P_{\mu}\cos\theta_{\gamma})
(d\Omega_{\gamma}/4\pi)$, where $f_{e}$ and $f_{\gamma}$ are an
integrated fraction of the spectrum of $e^{+}$ in the normal muon
decay and that of photon in $\mu^{+}\rightarrow e^{+}\nu
\overline{\nu}\gamma$ decay within the signal region,
respectively. $f^{0}_{e}$ and $f^{0}_{\gamma}$ include their
corresponding branching ratios. Here, for simplicity, the source of a
high energy photon is assumed to come mostly from the radiative muon
decay. $\Delta t$ and $\Delta \omega$ are respectively a full width of
the signal regions for timing coincidence and angular constraint of
the back-to-back kinematics.  $\eta$ is a suppression factor of the
accidental background. When $\eta$ is small, the accidental background
is suppressed. $\eta = 1$ for the case of unpolarized muons.

Given the sizes of the signal region, $B_{acc}$ can be estimated. Take
$\delta x$, $\delta y$, $\delta \theta_{e\gamma}$, $\delta t$ to be a
half width of the signal region for $e^{+}$ and photon energies, angle
$\theta_{e\gamma}$ and relative timing between $e^{+}$ and photon,
respectively.  $f^{0}_{e}$ can be estimated by integrating the Michel
spectrum of the normal muon decay from $1-\delta x \leq x \leq 1+r$,
yielding $f^{0}_{e} \approx 2(\delta x)$. $f^{0}_{\gamma}$ can be
given from Eq.(\ref{eq:int50}). $\Delta \omega/4\pi$ is $(\delta
\theta_{e\gamma})^2/4$.  For instance, to obtain quantitative
estimation, take some reference numbers such as $\delta x = 0.005$,
$\delta y = 0.03$, $\delta \theta_{e\gamma} = 0.01$ radian, $\delta t$
= 0.5 nsec, and $R_{\mu} = 3 \times 10^{8}~\mu^{+}$/sec, $B_{acc}$ is
$3 \times 10^{-13}$. Unless there are significant improvements made on
the detector resolution, the accidental background might appear at a
level of $10^{-13}$.

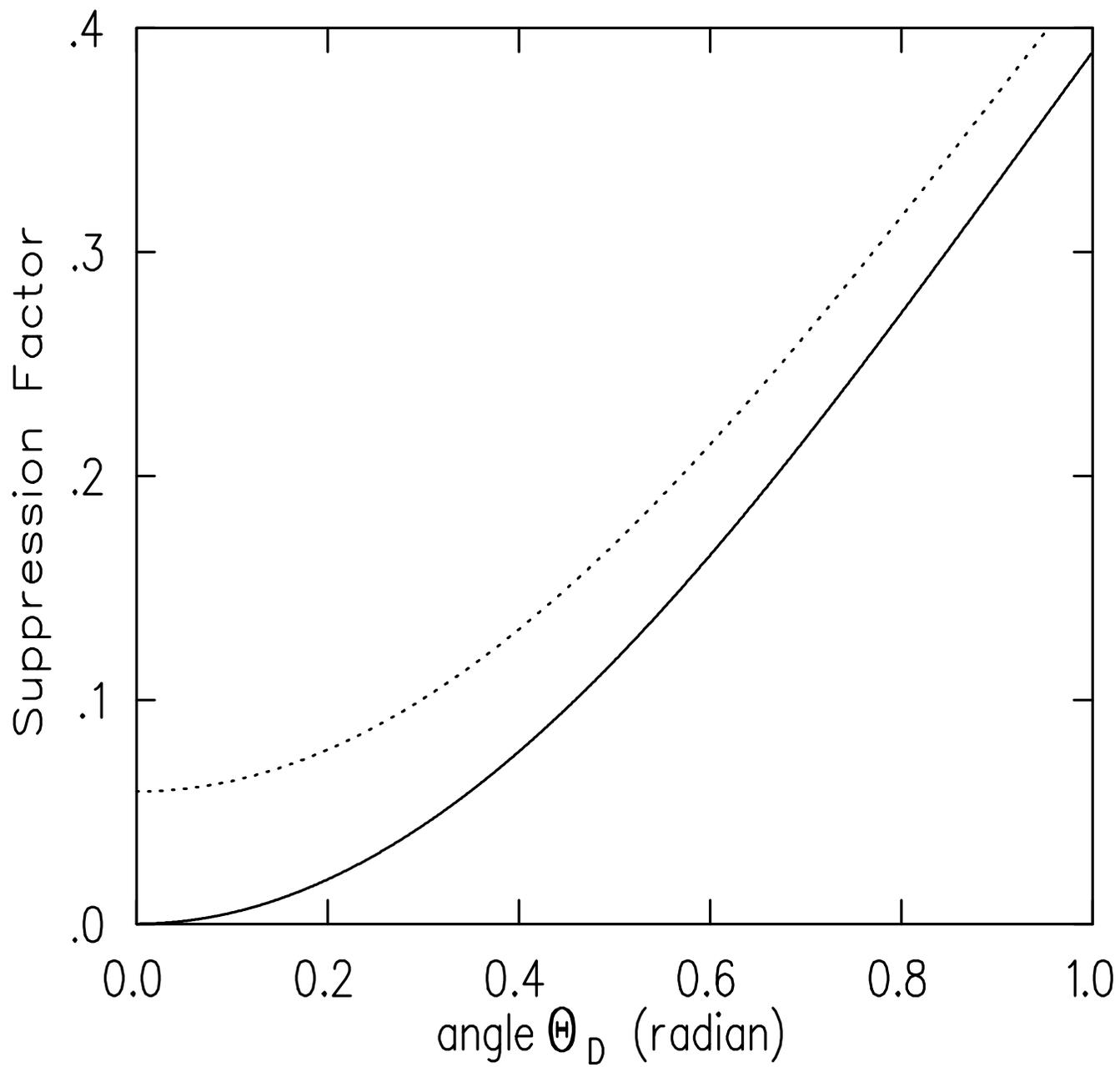
\begin{figure}
\caption{Suppression factor $\eta$ of accidental backgrounds as a
function of a half of the opening angle of detection ($\theta_{D})$ (a
solid line for $P_{\mu}$ = 100 \% and a dotted line for $P_{\mu}$ = 97
\%).}
\label{fg:cos}
\end{figure}

With the use of polarized muons, however, the accidental background
can be suppressed further. The selective measurement of either
$e^{+}$s or photons antiparallel to the muon spin direction with a
medium solid angle would provide a large background suppression
factor.  By taking account of the angular distributions of $e^{+}$s
and photons which are supposed to be back-to-back for the $\mu^{+}
\rightarrow e^{+} \gamma$ signal, the suppression factor $\eta$ is
given for polarized muons by

\begin{eqnarray}
\eta &\equiv& \int^{1}_{\cos\theta_{D}}d(\cos\theta) (1 +
P_{\mu}\cos\theta)(1 - P_{\mu}\cos\theta) /
\int^{1}_{\cos\theta_{D}}d(cos\theta) \cr 
&=& (1-P_{\mu}^2) + {1\over3}P_{\mu}^2(1-\cos\theta_{D})(2+\cos\theta_{D})
\label{eq:cos}
\end{eqnarray}

\noindent where $\theta_{D}$ is a half opening angle of detection with
respect to the muon polarization direction. In Fig.\ref{fg:cos}, the
suppression factor of the accidental background is shown as a function
of $\theta_{D}$. To obtain a high suppression factor, the detector
solid angle has to be adequately modest. For instance, for $\theta_{D}
= 300$ mrad, an accidental background can be suppressed down to about
$1/20$ when $P_{\mu}$ is 100 \%.

As seen in Eq.(\ref{eq:cos}), however, if $P_{\mu}$ is not
sufficiently high, $\eta$ might be dominated by $1 - P_{\mu}^2$, and
beyond it no further suppression can be obtained even if making
$\theta_{D}$ smaller. Therefore, it is inevitable to have $P_{\mu}$ as
high as possible.  The previous experiment \cite{bolt88} used a
surface muon beam, which is muons from a decay of pions stopped near
the surface of the pion production target. From their production
mechanism, they are supposed to be 100 \% polarized opposite to the
muon momentum direction. In reality, the beam channel acceptance and
multiple scattering in the production target will decrease the muon
polarization, but it should be arranged to be at least 97 \% or more.
For instance, a muon polarization of 97 \% will give us a suppression
of about $\eta = 1/10$ for $\theta_{D} = 300$ mrad, as seen in
Fig.\ref{fg:cos}. Similarly, a deviation from a pure $1+\cos\theta$
distribution of $e^{+}$s and photons would introduce the same effects.
For instance, for $\delta y$ = 0.03, the residual at
$\cos\theta_{\gamma} = -1$ changes a suppression factor from 1/20 to
1/8 for $\mu^{+} \rightarrow e^{+}_{L} \gamma$ with $\theta_D$ = 300
mrad and $P_{\mu}$ = 100 \%. This would be improved if $\delta y$
becomes smaller. Since the $e^{+}$ energy resolution is better than
the photon energy resolution, this effect is much smaller for $\mu^{+}
\rightarrow e^{+}_{R} \gamma$.

An opening angle of the detector of $\theta_{D}$ = 300 mrad, taken as
an example in this paper, gives a geometrical solid-angle coverage of
2.3 \% for each side. If the $\mu^{+} \rightarrow e^{+} \gamma$
signals are preferentially emitted, following a $1 \pm
P_{\mu}\cos\theta$ distribution, towards the detector coverage, the
effective signal acceptance could be increased further by a factor of
two, resulting in the additional improvement of the
signal-to-background ratio by a factor of two. This solid angle would
be large enough to yield a single event sensitivity to
$\mu^{+}\rightarrow e^{+} \gamma$ at the level of $10^{-14}$ or less
for the muon beam intensity presently available, together with
sufficient accidental background suppression thus discussed. It should
be noted that the physics background from $\mu^{+} \rightarrow e^{+}
\nu \overline{\nu} \gamma$ decay is also estimated to be less than
$10^{-14}$ for the detector resolutions given in this paper
\cite{kuno96}.

In conclusion, with polarized muons, a search for both $\mu^{+}
\rightarrow e^{+}_{L} \gamma$ as well as $\mu^{+} \rightarrow
e^{+}_{R} \gamma$ has the potential for improved sensitivity in terms
of accidental background suppression. The suppression in the search
for $\mu^{+} \rightarrow e^{+}_{R}\gamma$ decay comes from the angular
distribution of $e^{+}$s in the normal muon decay, whereas that for
$\mu^{+} \rightarrow e^{+}_{L} \gamma$ decay is due to the
distribution of high energy photon in $\mu^{+} \rightarrow e^{+} \nu
\overline{\nu} \gamma$ decay. With a reasonable detector acceptance, a
rejection of an order of magnitude could be achieved.

\acknowledgments The authors are acknowledged for Drs.  A. Van der
Schaaf and H.K. Walter for their useful discussions. The authors of
Y.K. and Y.O. are grateful for the hospitality at the Institute for
Theoretical Physics, University of California Santa Barbara during
their stays.  This work was supported in part by the Grant-in-Aid of
the Ministry of Education, Science, Sports and Culture, Government of
Japan, and in part by the National Science Foundation under Grant
No. PHYS94-07194.

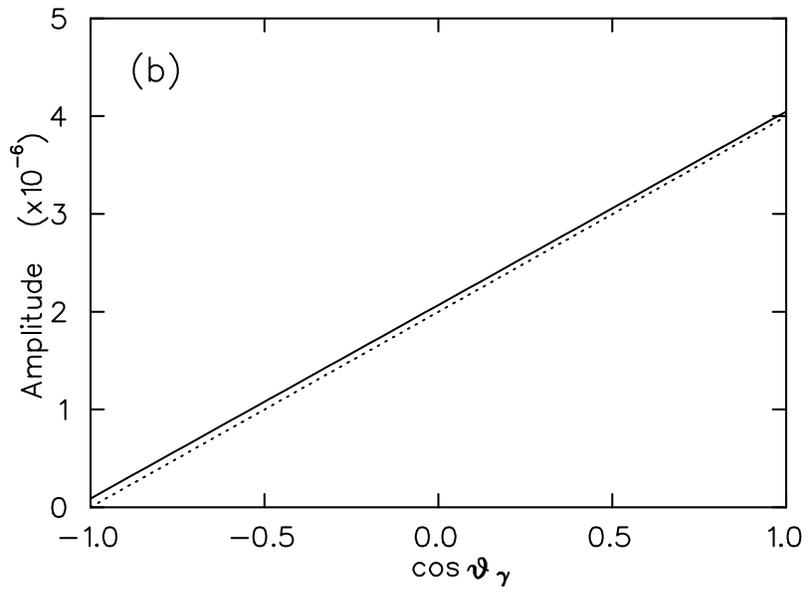
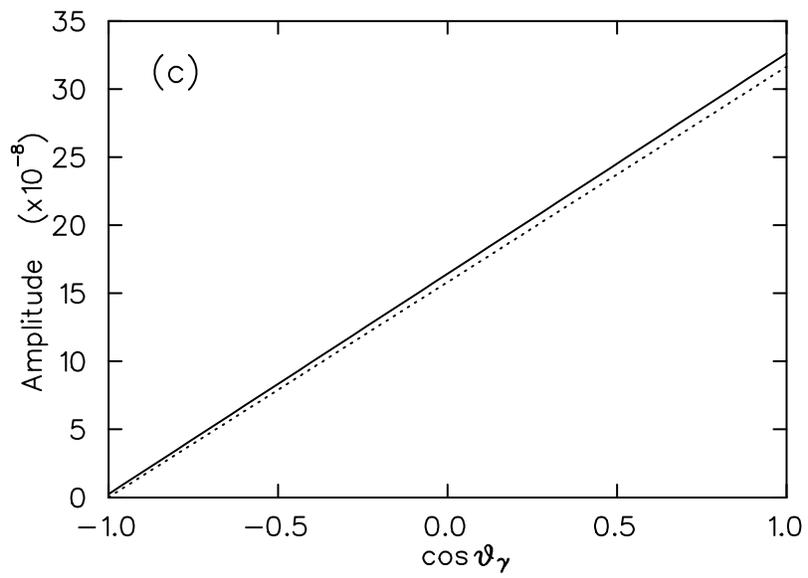
\end{document}